# Nonreciprocally Boosting Magnetoacoustic Coupling with Surface-Acoustic-Wave-induced Spin Transfer Torque

**Shuting Cui[1#], Fa Chen[1#], Liyang Liao[2], Jiacheng Lu[1], Rui Xiong[3], Xiaofei Yang[1], Yoshichika Otani[2, 4, 5], Yue Zhang[1, 6*], Wei Luo[1*]**

**[1]School of Integrated Circuits, Huazhong University of Science and Technology, Wuhan, 430074, China**

**[2]Institute for Solid State Physics, University of tokyo, Kashiwa 277-8581, Japan.**

**[3]School of Physics and Technology, Wuhan University, Wuhan 430072, China**

**[4]Center for emergent Matter Science, RiKen, Wako, Saitama 351-0198, Japan.**

**[5]Trans-scale Quantum Science institute, University of tokyo, tokyo 113-8654, Japan.**

**[6]Songshan Lake Materials Laboratory, Dongguan, Guangdong 523808, China**

**#Shuting Cui and Fa Chen contributed equally to this work.**

***Corresponding author: yue-zhang@hust.edu.cn (Yue Zhang); luowei@hust.edu.cn (Wei Luo)**

Strengthening magnetoacoustic coupling is crucial to the improvement of the surface acoustic wave (SAW)-driven spintronics devices. A key challenge in enhancing magnetoacoustic coupling is minimizing the phonon and magnon dissipation of the device, which usually requires complicated techniques for generating shear-horizontal (SH) or standing waves to suppress the phonon dissipation. In this work, we significantly strengthened the magnetoacoustic coupling by suppressing the magnon dissipation via the SAW-induced spin-transfer-torque (STT) in Co/Cu/NiFe multilayer, which is facilitated by the non-parallel magnetization alignment between the two ferromagnetic layers. Also, this STT exhibits the form of Zhang-Li torque due to the SAW-induced spin wave, which gives rise to the unique nonreciprocal SAW transportation under external magnetic field. This finding opens new avenues for non-reciprocally boosting magnetoacoustic coupling, which pays the way for developing on-chip SAW-driven multifunctional devices.

## INTRODUCTION

In recent years, surface acoustic waves (SAWs) have emerged as an attractive route for triggering magnetic dynamics, such as ferromagnetic resonance, spin current, nonreciprocal magnetoelastic waves, and magnetization switching[1–4]. Thanks to the advantage in energy efficiency, SAW-driven magnetoacoustic devices are highly promising for applications in magnetic field sensing, radio-frequency isolators and circulators, and information processing[5–8].

The strength of magnetoacoustic coupling relies on the magnetoelastic coupling and the dissipation from magnon and phonon. In addition to enhancing magnetoelastic interaction, suppressing dissipation is also crucial to the strengthening of magnetoacoustic coupling[9]. One recent approach is to construct high-Q acoustic resonators to generate SH-type standing wave to suppress phonon loss[10], which requries stringent microfabrication techniques.

To enhance magnetoacoustic coupling, reducing magnon dissipation is the same important to depressing phonon dissipation[11]. Up to now, it is widely believed that controlling magnon loss is experimentally challenging, since spin current or spin-polarized current injected by external electrical current is usually necessary for inhibiting magnetic damping torque[12]. However, in addition to the injection of electrical current, the SAW itself can generate spin currents in Ferromagnet/Nonmagnet (FM/NM) multilayers[4], but these SAW-induced spin currents have not been actively utilized to control the magnon dissipation and enhance magnetoacoustic coupling.

In this study, we demonstrate that in a Co/Cu/NiFe trilayer on a $LiNbO_3$ substrate, exploiting the SAW-induced spin current can significantly enhance the magnetoacoustic coupling based on the spin-transfer-torque (STT) effect. The basic physical process for the enhancement of magnetoacoustic coupling by SAW-induced STT was illustrated in **Fig. 1**. Initially, the Rayleigh-type SAW triggers weak magnetization precession in the bottom Co layer. Based on the spin-pumping effect, the magnetization precession in Co generates spin current that traverses the intermediate Cu layer to interact with the magnetization in the upper NiFe layer [**Fig. 1(a)**]. Here Cu is famous for its long diffusion length of spin current[13,14]. When the magnetizations of Co and NiFe layers are non-parallel, this spin current generates STT that drives the magnetization precession in the upper NiFe, which, in turn, produces a backward spin current with STT that strenthens the magnetization precession in the bottom Co layer and enhances the SAW absorption [**Fig. 1(b)**].

This work also discovers that the STT exhibits the form of Zhang-Li torque due to the SAW-induced spin wave, and the sign of STT relies on the direction for external magnetic fields or SAW transportation. This approach paves the way for non-reciprocally enhancing magnetoacoustic coupling based on the SAW-induced STT, which can be applicable for on-chip magnetoacoustic devices with unidirectional SAW transportation under magnetic field.

## RESULTS

### 1. Magnetic-field-dependent $S_{21}$ and magnetic hysteresis loops (M-H)

The device structure is shown in **Fig. 2(a)**. The interdigital transducers (IDTs) made of Al were fabricated on the surface of the $LiNbO_3$ single-crustal substrate to generate Rayleigh-type SAWs by a vector network analyzer (VNA) connected to the IDTs. The forward transmission coefficient ($S_{21}$) spectrum exhibits the maximum value near 1.6 GHz and 2.7 GHz. In the intermediate region between the two IDTs, the Co/Cu/NiFe thin film was deposited by using direct current (DC) magnetron sputtering. Both of the Co and NiFe have a thickness of 20 nm. The thickness of the Cu layer is 8 nm or less. In this thickness range of Cu, the spin current from spin-rotation coupling in Cu can be safely disregarded[14]. The variation of the forward transmission parameter $S_{21}$ was measured as a function of in-plane magnetic field strength (***H***) along different orientations.

**Fig. 2(b)** demonstrates the relationship between $S_{21}$ for the 1.6-GHz SAW and magnetic field strength for four structures: Co(20)/Cu(8)/NiFe(20)/Pt(5), Co(20)/NiFe(20)/Pt(5), Co(20)/Cu(8)/Pt(5), and NiFe(20)/Cu(8)/Co(20)/Pt(5) under magnetic field orientations of $\varphi_H = \pi/2$ (perpendicular to the SAW propagation). The results reveal that the change of $S_{21}$ of Co(20)/Cu(8)/NiFe(20)/Pt(5) was significantly enhanced. However, in the other three structures, the $S_{21}$ change remains weak at all magnetic field orientations. We measured the $S_{21}$ along different magnetic-field orientations and calculated the maximum absolute value of the changing of $S_{21}$ ($|\Delta S_{21}|_{max}$). As shown in **Fig. 2(c)**, the $|\Delta S_{21}|_{max}$ at different magnetic field angles verified the most significant absorption of SAW for Co(20)/Cu(8)/NiFe(20)/Pt(5) at $\varphi_H = \pi/2$ and $\varphi_H = 3\pi/2$. This is different from the typical angular dependence of magnetoelastic coupling, which satisfies $\sin^2(2\varphi_H)$ with the strongest SAW absorbtion at $\varphi_H = (n + 1/2)\pi/2$ with $n$ = 0, 1, 2, and 3[15,16]. All the other three samples exhibit much weaker $|\Delta S_{21}|_{max}$ along different directions.

The comparison of the results in **Fig.2 (b) ~ (c)** indicates that the intersection of Cu between Co and NiFe plays a crucial role in the significant enhancement of SAW absorption. In the case of magnetoacoustic coupling involving only the Co layer, even though the magnetoelastic coupling coefficient of Co is close to that of Ni[17], the SAW absorption for Co is extremely weaker, with a value of just 0.02 dB. This is due to the high magnon dissipation of Co with a large damping coefficient. Additionally, the SAW absorption is also negligible when the sequence of NiFe and Co is exchanged. In this case, the magnetoelastic coupling constant of the bottom NiFe is negligible[18]. Therefore, the combined effects of spin current and magnetoacoustic coupling is necessary for the significant enhancement of SAW absorption. Since the Ruderman–Kittel–Kasuya–Yosida (RKKY) coupling between the two FM layers is generally negligible due to the Cu layer being significantly thicker than 2 nm in this paper, the SAW-triggered magnetization dynamics observed in this study fundamentally differ from the recently reported magnetic dynamics induced by RKKY coupling[16].

To account for the mechanism of the strengthening the magnetoacoustic coupling, we measured the magnetic hysteresis loops (*M-H*) of the Co(20)/Cu(8)/Pt(5) and Cu(8)/NiFe(20)/Pt(5) systems by using the vibrating sample magnetometer (VSM), both of which exhibit clear uniaxial anisotropy [**Figs. 2(d) and (e)**]. The easy axes of the NiFe and Co layers were oriented at a π/2 angle to each other. These differences in magnetic properties between NiFe and Co lead to a non-collinear alignment of the magnetization of the two FM layers under a small applied field, with an angle $\varphi_H$ between 0 and π/2. This strengthened the STT effect that enhanced the SAW absorption. At $\varphi_H = 0$, the SAW absorption due to magnetoelastic coupling is weaker than at $\varphi_H = \pi/2$ as the easy axis of Co is along the direction for SAW transportation [**Fig. 2(c)**]. Consequently, the spin-current-induced enhancement of SAW absorption at $\varphi_H = 0$ is significantly weaker than at $\varphi_H = \pi/2$.

To further verify the role of spin current, we replaced Cu with Pt of varying thickness in the middle layer on $S_{21}$ (The total thickness of Cu/Pt/Cu layer remained constant at 8 nm.). **Fig. 3 (a)** demonstrates that the interlayer Pt significantly attenuates variations in the change of $S_{21}$. This effect arises from the obstruction of spin current transmission in Pt, since the spin diffusion length of Cu is significantly larger than that of Pt [**Fig. 3 (b)**]. This finding further confirms that the spin-current transmission between Co and NiFe plays a crucial role in enhancing $S_{21}$.

**2. Nonreciprocal high-frequency SAW transportation induced by Zhang-Li torque**

The $S_{21}$ for Co(20)/Cu(8)/NiFe(20)/Pt(5) in **Fig. 2(b)** exhibits weak asymmetry with opposite directions of external magnetic field or SAW transportation. Up to now, the nonreciprocal SAW transportation has been observed in magnetoacoustic coupling devices with interfacial DMI, interlayer RKKY coupling, magneto rotational coupling, etc[16,19,20]. In Co(20)/Cu(8)/NiFe(20)/Pt(5), however, all of these factors are negligible due to the large thickness of Co, NiFe, Cu, and the in-plane magnetic anisotropy. We show that the presence of spin current and STT offers unique mechanism for the nonreciprocity.

Since the spin-current strength and the STT from spin pumping is proportional to frequency, to unveil the mechanism of the nonreciprocity, we measured the backward and forward transmission parameters $S_{12}$ and $S_{21}$ in the Co(20)/Cu(8)/NiFe (20)/Pt(5) structure at a higher frequency (2.7 GHz) as a function of the external magnetic field. We clearly observed asymmetric $S_{21}$ under positive and negative magnetic fields. At $\varphi_H = \pi/2$, the propagation is characterized by a single-polarity, single-peak response [**Fig. 4(a)**]. Effectively attenuates SAW propagating in one magnetic field direction while leaving propagation in the opposite direction almost unaffected. At $\varphi_H = 5\pi/12$, a distinct asymmetry also emerges, where a single peak appears under one field polarity, while a double-peak was observed under the opposite polarity [**Fig. 4(b)**]. Also, the asymmetric $S_{21}$ is accompanied by nonreciprocal SAW transportation, i.e., under the magnetic field with a fixed direction, the SAW can pass with negligible attenuation along one direction, but the SAW absorption was significantly enhanced along the opposition orientation for transportation.

First, we discuss about the asymmetry for the SAW absorption under magnetic field with opposite directions. In conjunction with **Fig. 4**, it can be determined that the spin current exerts a STT which is relevant to the angle of magnetization between NiFe and Co. Since the two FM layers exhibit distinct coercivity and uniaxial anisotropy fields, the magnetization reversing in the two FM layers is not synchronous. In the initial reversal process, the magnetization of Co starts the rotation before NiFe due to its easy axis along the *x*-axis direction. which enhances the STT with the increase of the angle of the magnetization between NiFe and Co from 0 to 90 degrees [**Fig. 4 (c)**]. However, when the direction of external magnetic is reversed, the magnetization in NiFe abruptly switched, which reversed the direction of STT and significantly depressed the SAW absorption. At a higher reversing magnetic field, the angle between NiFe and Co reduces to be acute degrees with gradual rotation of the magnetization of Co. However, the intrinsic damping torque also enhances due to the stronger magnetic field, which also depressed the SAW absorption [**Fig. 4 (d)**]. At $\varphi_H = 5\pi/12$, this nonreciprocity is much clearer, since the magnetic field is neither parallel nor perpendicular to the easy axis, and the magnetization switching in either NiFe or Co is not too sharp.

To understand the nonreciprocal SAW transportation, we propose that the STT exhibits Zhang-Li torque by taking into account the spin wave propagation driven by SAW, since the length of the film between the two IDTs (375 μm) is much longer than the SAW length (1.44 μm for 2.7 GHz SAW). Therefore, the damping-like STT has the form $c\mathbf{m}\times\left(\mathbf{m}\times\frac{\partial \mathbf{m}'}{\partial x}\right)$ with **m** and **m'** the unit magnetization in the two FM layers[21]. For the spin wave excited by SAW, the **m** and **m'** can be expressed as $\mathbf{m}(\mathbf{m}')\sim e^{\mathrm{j}(kx-\omega t)}$, which converted the STT expression as $\sim\ ck\mathbf{m}\times(\mathbf{m}\times\mathbf{m}')$. The sign of this term changed while reversing the direction for the SAW and spin-wave propagation, which explains the appearance of the nonreciprocity observed in **Fig. 4**.

### 3. Calculation of absorption power under the SAW-induced STT

In order to theoretically verify the role of spin current and STT in enhancing the magnetoacoustic coupling, we calculated the spin dynamics in the Co/Cu/NiFe trilayer based on the Landau-Lifshith-Gilbert (LLG) equations with the damping-like torque from the spin current:

$$\frac{\partial \mathbf{m}}{\partial t} = -\gamma \mathbf{m}\times\mathbf{H}_{\mathbf{eff}} + \alpha \mathbf{m}\times\frac{\partial \mathbf{m}}{\partial t} + \mathrm{j}ck\mathbf{m}\times\left(\mathbf{m}\times\mathbf{m}'\right) \tag{1}$$

$$\frac{\partial \mathbf{m}'}{\partial t} = -\gamma' \mathbf{m}'\times\mathbf{H}'_{\mathbf{eff}} + \alpha' \mathbf{m}'\times\frac{\partial \mathbf{m}'}{\partial t} - \mathrm{j}c'k\mathbf{m}'\times\left(\mathbf{m}'\times\mathbf{m}\right) \tag{2}$$

where $\mathbf{m}$, $\mathbf{H}_{\mathbf{eff}}$, $\alpha$, $\gamma$, and $k$ are the magnetization, effective magnetic field, the Gilbert damping coefficient, gyromagnetic ratio, and the wave number of the upper NiFe layer, respectively. And these parameters of the bottom Co layer were labeled by the apostrophe in the top right corner. Here, $c$ is a coefficient characterizing the STT strength[22]. If the direction of the SAW transport is reversed, $k$ reverses its sign.

Based on the coupled LLG equation [**Eqs. (1) and (2)**], the absorption power of SAW can be expressed as $P_{\mathrm{abs}} \propto \int_0^{\lambda} \mathbf{h}^T \boldsymbol{\chi} \mathbf{h} dx$, where $\lambda$ is the wave length of SAW, $\boldsymbol{\chi}$ is the Polder susceptibility tensor, $\mathbf{h}$ represents the effective magnetic field including the contributions from STT and strain [**Supplementary Material I**].

When $\varphi_{\mathrm{H}} = \pi/2$, the absorption power of SAW by Co/Cu/NiFe was calculated. Our calculation results indicate that the interplay of the uniaxial anisotropic field and the magnetoelastic field causes little change in SAW energy in absence of the STT term ($c = 0$). When $c$ is nonzero, the magnitude of $ck$ or $c'k$ increases with the enhancement of the spin current, resulting in more SAW power being absorbed [**Fig. 5(a)**]. Furthermore, under the influence of the reverse magnetic field, the variation of SAW power exhibits asymmetry [**Fig. 5(b)**], due to the change in the STT direction as shown in **Fig. 4(c)-(e)**. When the magnetic field $H = +4.54$ mT (the magnetic field corresponding to the strongest absorption in **Fig. 5(b)**), the non-reciprocal transmission of the SAW can be obtained by changing the wave number $k$ of the SAW [**Fig. 5(c)**].

## SUMMARY

In summary, this study reveals that the SAW-induced spin current plays a crucial role in enhancing the coupling between SAW and magnetic precession in a Co/Cu/NiFe multilayer structure on a piezoelectric $LiNbO_3$ substrate. We demonstrate substantial contribution of spin current to the enhancement of SAW absorption, which is relevant to the STT due to the noncollinear alignment of the magnetization of the two FM layers. In addition to that, a non-reciprocal SAW transportation was observed based on the Zhang-Li torque originating from the SAW-induced spin wave and nonsynchronous reversing of magnetization in Co and NiFe. These findings not only deepen our understanding of the interaction between spin current and SAW but also open new avenues for the development of on-chip acoustic devices with the integration of multifunction based on the multiphysics coupling manipulation of SAW, magnetization, and spin current.

## MATERIALS AND MTHEODS

Using electron beam evaporation, an Al electrode was deposited on a 128°-YX single-crystal $LiNbO_3$ substrate. The thickness and width of the electrode was 72 nm and 600 nm, respectively. The distance between the neigboring electrodes was also 600 nm. Each IDT was made up of 25 pairs of electrodes, and the delay distance between two IDTs was 1440 μm. Samples were fabricated on the $LiNbO_3$ substrates by using high vacuum DC magnetron sputtering. A hard mask was employed to define the magnetic film with the size 375 μm × 1500 μm, and the Co(20)/Cu(8)/NiFe(20)/Pt(5) (numbers in the parentheses are thicknesses in nanometres) multilayers were deposited under a base pressure better than $1 \times 10^{-7}$ torr. Ar gas (purity > 99.999 %) was ultilized in the sputtering. Co, Cu, NiFe and Pt layers were deposited at an Ar pressure of 0.3 Pa with the sputtering rate of 25.1 s/nm, 8.2 s/nm, 16 s/nm, and 7.6 s/nm, respectively. The 99.99% $Ni_{81}Fe_{19}$ target material sputtered to form NiFe thin layer. A vector network analyzer (Keysight E5071C) was exploited to inject and detect the SAW signals. An electromagnet was placed under the sample to provide an in-plane magnetic field. The magnetic properties such as *M-H* curves were measured by a vibrating sample magnetometer equipped in a Physical Property Measurement System (Quantum Design).

## FIGURES

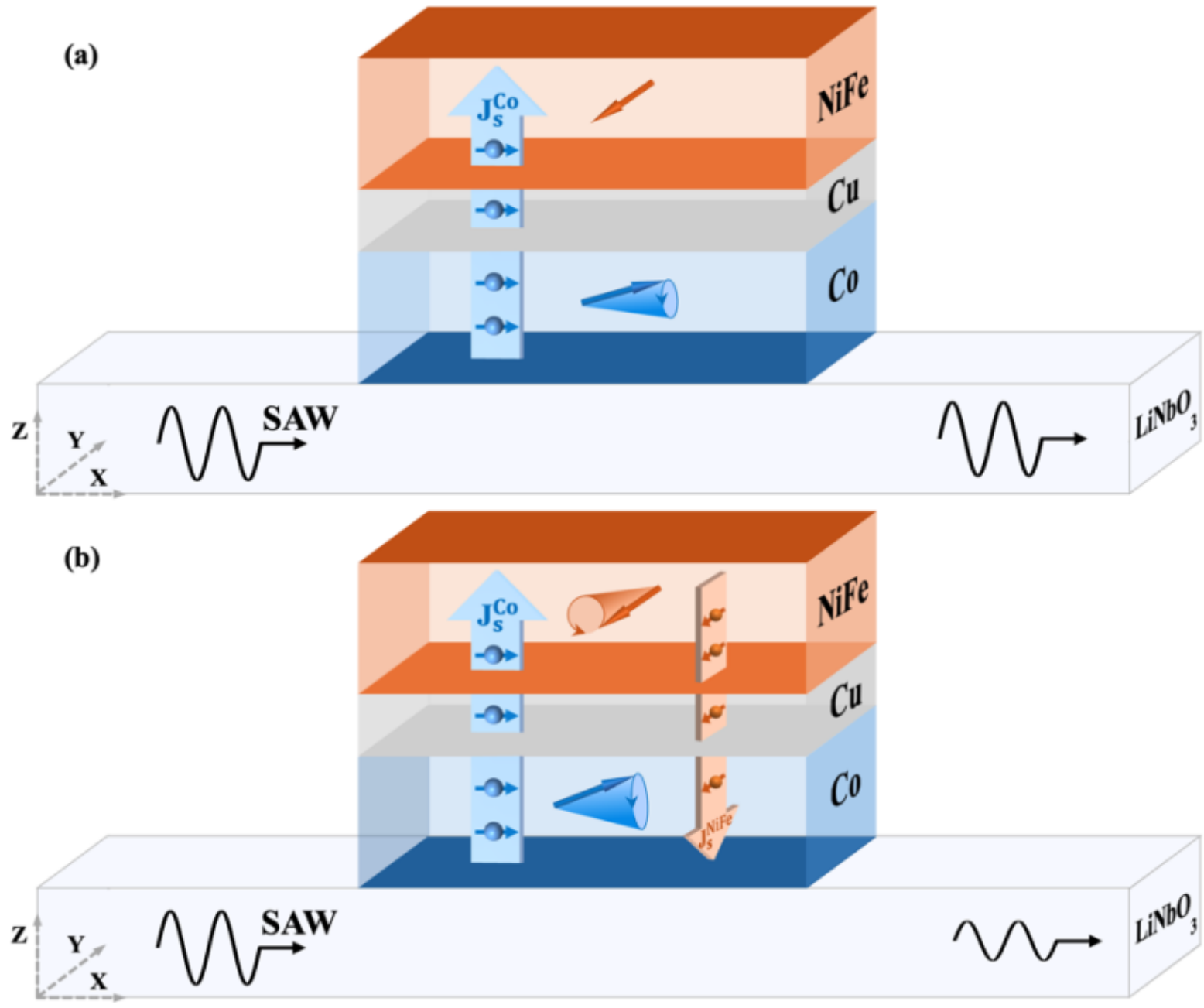

**Figure. 1 Illustration of the mechanism for the enhancement of magnetoacoustic coupling by the SAW induced spin current. (a) The SAW drives the magnetization precession in the bottom Co layer, generating spin currents that propagate upward. (b) These spin currents drive the magnetization precession in the upper NiFe layer based on STT and, in turn, generate a reverse spin current downward, enhancing the bottom magnetization precession and significant enhancement of the absorption of the SAW.**

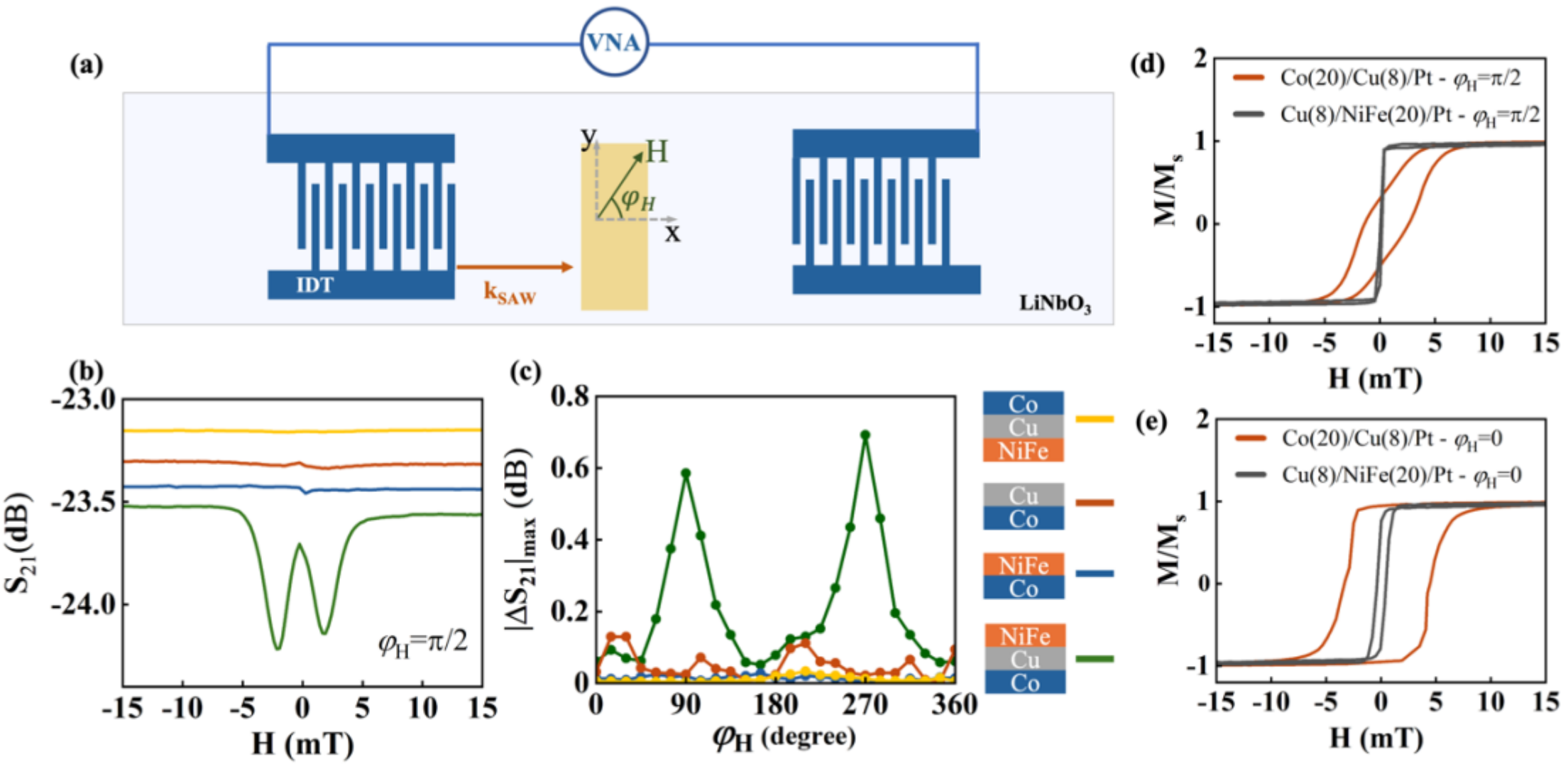

**Figure. 2 (a) Schematic of the device, featuring a pair of IDTs on a $LiNbO_3$ substrate, with the film deposited near the center of the delay line. The measurements were performed by using a VNA. (b) Dependence of $S_{21}$ on the external magnetic field $H$ at $\varphi_H = \pi/2$. (c) Maximum variation of $S_{21}$, $|\Delta S_{21}|_{max}$, as a function of the orientations of external magnetic fields. Both (b) and (c) are shown for the following film structures: NiFe(20)/Cu(8)/Co(20)/Pt(5) (yellow line), Co(20)/Cu(8)/Pt(5) (red line), Co(20)/NiFe(20)/Pt(5) (blue line) and**

Co(20)/Cu(8)/NiFe(20)/Pt(5) (green line). (d) and (e) show the *M-H* curves for Co(20)/Cu(8)/Pt(5) and Cu(8)/NiFe(20)/Pt(5) at $\varphi_H = 0$ and $\varphi_H = \pi/2$.

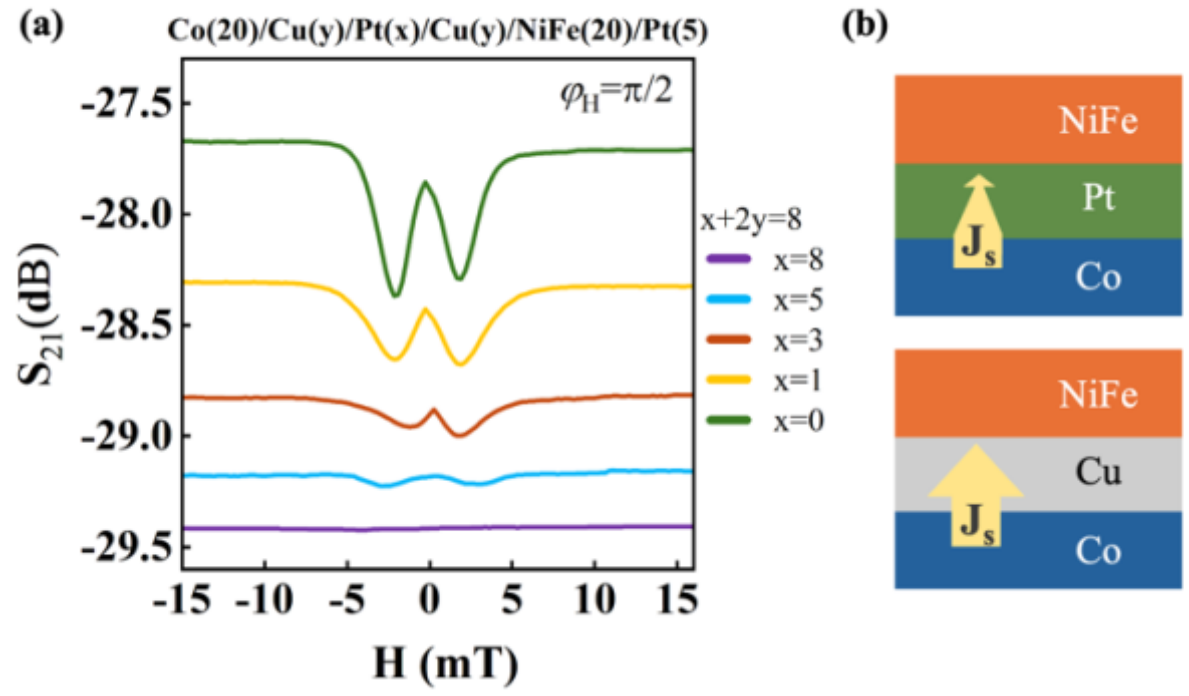

Figure. 3 (a) $S_{21}$ with the Cu spacer layer replaced by Pt at varying thicknesses for $\varphi_H = \pi/2$. (b) Schematic illustration of spin current propagation through two non-magnetic layers, Cu and Pt.

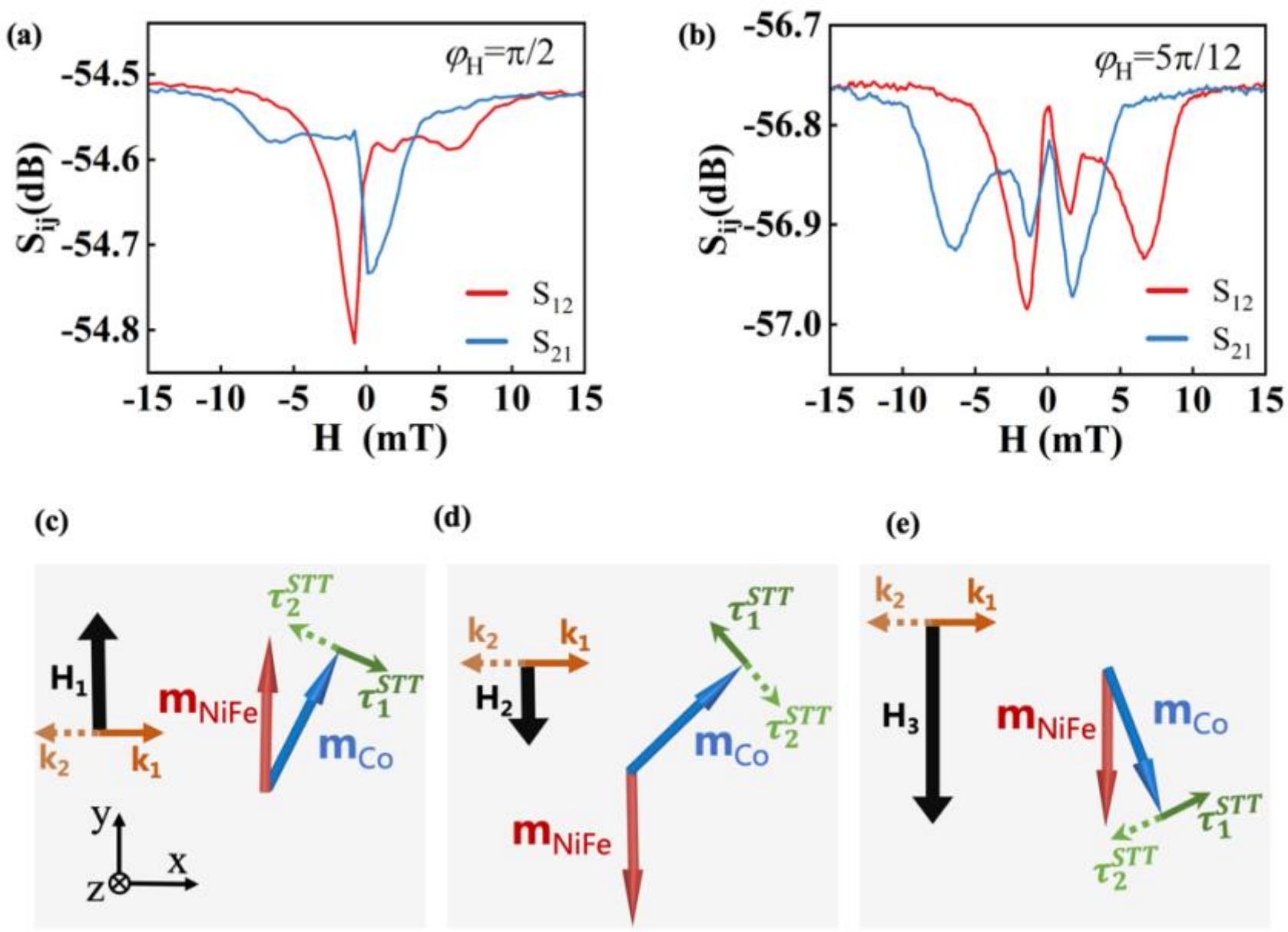

Figure. 4 (a) Transmission parameter $S_{12}$ and $S_{21}$ at 2.7 GHz for $\varphi_H = \pi/2$ and $\varphi_H = 5\pi/12$. (b) Nonsynchronous reversing of magnetization of Co and NiFe and the STT with the reversing of magnetic field $\vec{H}$ and wave number $\vec{k}$ for SAW.

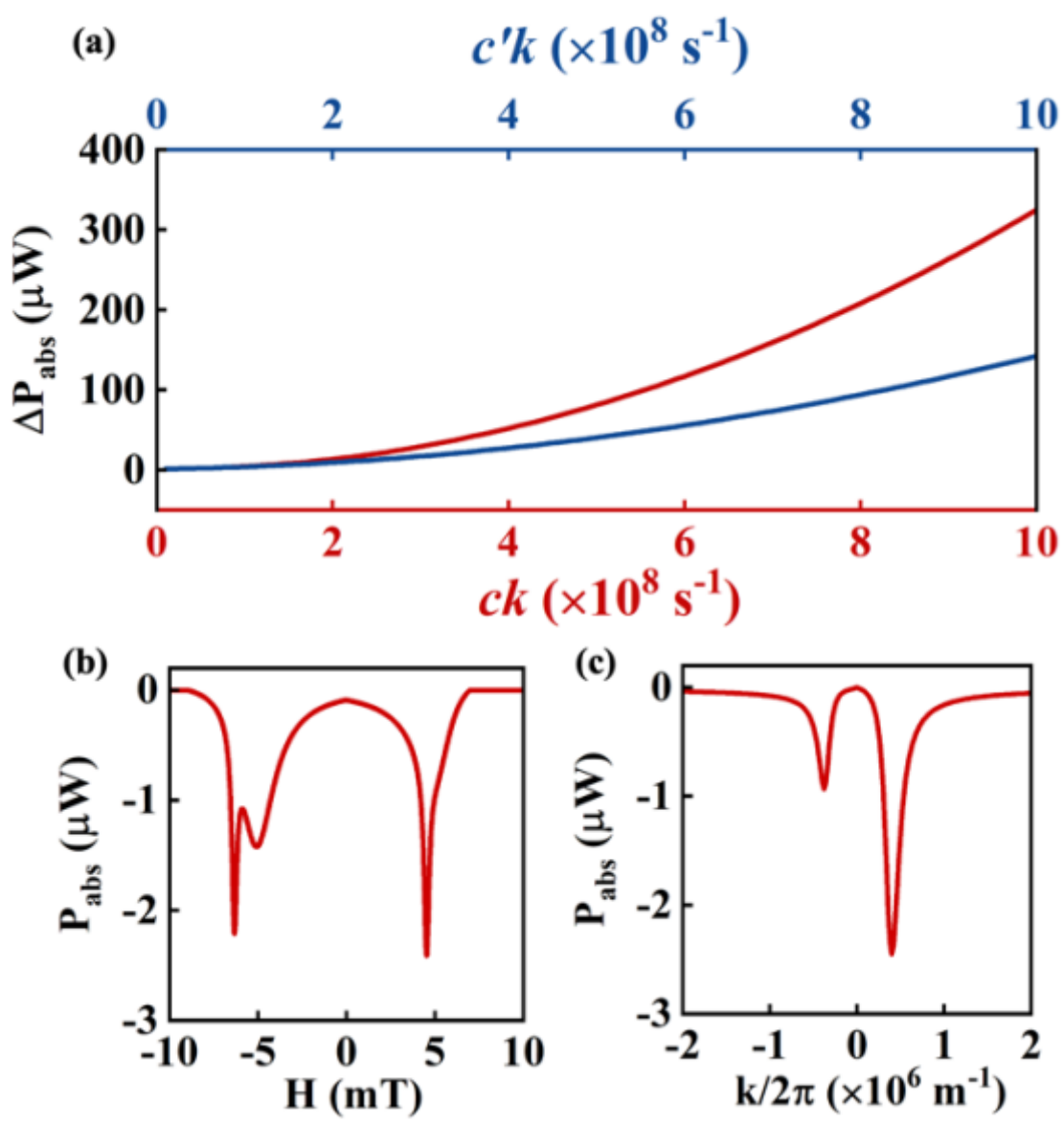

**Figure. 5 (a) Calculated absorption power of SAW varies with the *ck* (red line) and *c'k* (blue line) from NiFe and Co. The absorption power as a function of magnetic field (b) and wave number (c) for $ck = c'k = 5 \times 10^7$ $s^{-1}$.**

## Acknowledgements:

This work was supported by the National Key Research and Development Program of China (Grant No. 2022YFE0103300), and the National Natural Science Foundation of China (Grant No. U2141236 and 12227806), and the open research fund of Songshan Lake Materials Laboratory (Grant No. 2023SLABFN26).